\documentclass[report]{seismica}

\title{SeisBench DAS: A machine learning framework for Distributed Acoustic Sensing}

\reporttype{Software Report}

\author[1]{Jannes~Münchmeyer
	\orcid{0000-0002-4006-9673}
	\thanks{Corresponding author: munchmej@gfz.de}
}
\author[1]{Han~Xiao
	\orcid{0000-0001-8727-8053}
}
\author[1,2]{Frederik~Tilmann
	\orcid{0000-0002-7439-8782}
}
\affil[1]{GFZ Helmholtz Centre for Geosciences}
\affil[2]{Institute for Geological Sciences, Freie Universität Berlin, Berlin, Germany}

\credit{Conceptualization}{Jannes Münchmeyer, Han Xiao, Frederik Tilmann}
\credit{Software}{Jannes Münchmeyer, Han Xiao}
\credit{Writing - Original draft}{Jannes Münchmeyer}
\credit{Writing - Review \& Editing}{Jannes Münchmeyer, Han Xiao, Frederik Tilmann}
\credit{Funding acquisition}{Frederik Tilmann}

\begin{document}

\makeseistitle{
	\begin{summary}{Abstract}
	Fibre optic sensing, such as distributed acoustic sensing (DAS), has become a widespread technology for geophysical studies.
	To process the large-scale datasets produced by DAS, several machine learning methods have been proposed.
	However, without standardization of data and models, these methods lack comparability and interoperability.
	This introduces a gap between model developers and practitioners analyzing DAS data and inhibits adoption of deep learning for DAS.
	To address these limitations, here we present SeisBench DAS, an extension to the SeisBench library for machine learning in seismology.
	SeisBench DAS defines standard formats for DAS benchmark datasets, including standardised metadata and labels, and DAS models.
	It builds on the xdas framework for data ingestion and virtual array handling, and on PyTorch for reading and applying the machine learning models.
	Importantly, SeisBench provides an engine to efficiently apply deep learning models to diverse formats of DAS data, bridging the gap between model developers and practitioners.
	SeisBench DAS is designed as an open and extensible framework, allowing to easily incorporate future developments in deep learning for DAS.
	\end{summary}
	\begin{summary}{Non-technical summary}
	In recent years, technologies for fibre optic sensing have become widely available.
	These technologies, such as distributed acoustic sensing (DAS), turn an optical fibre into a dense array of seismic sensors.
	As fibres are comparatively cheap to deploy and recordings can also be conducted on already deployed telecommunication fibres, this allows instrumenting regions such as volcanoes, glaciers, and ocean bottoms.
	However, with the dense spacing of sensors, DAS also produces thousands of times more data than classical instruments.
	To effectively process this wealth of data, different machine learning methods have been proposed.
	However, these methods lack standardization and expose a gap between model developers and seimological practitioners.
	Here, we introduce SeisBench DAS, a library for processing DAS data.
	SeisBench DAS provides standardized interfaces for models and datasets, a benchmark data format specification, and an engine to easily apply deep learning models to diverse formats of DAS data.
	With an open and extensible design, SeisBench DAS aims to build a community framework for machine learning on DAS data.
	\end{summary}
}%

\section{Introduction}

Over the last decade, fibre-optic sensing has revolutionized observational seismicity \citep{sladenDistributedSensingEarthquakes2019,walterDistributedAcousticSensing2020,joussetFibreOpticDistributed2022,sladenDistributedSensingEarthquakes2019}.
Novel techniques, such as distributed acoustic sensing (DAS), have lead to an unprecedented growth in the density of observations.
In addition, fibre-optic sensing makes continuous monitoring of previously inaccessible regions feasible and financially viable, such as the ocean bottom, active volcanoes, or glaciers \citep{walterDistributedAcousticSensing2020,joussetFibreOpticDistributed2022,xiaoFrequencyDependentMicroseisms2025,liMinutescaleDynamicsRecurrent2025}.
At the same time, DAS data introduces a host of novel challenges, among which one of the foremost is the sheer volume of data \citep{wangDistributedAcousticSensing2025}.
For example, at 100 Hz sampling rate and with 32 bit precision, a fibre with 10,000 channels already produces around 4 MB/s, comparable to about 3 hours of classical, uncompressed, single-channel seismic data.
This explosion of data, together with the physical particularities of DAS data, such as azimuthal sensitivity variations and inconsistent coupling, requires the development of novel methods for observational seismology.

A promising approach to DAS processing is machine learning.
In particular, deep learning is optimally suited for DAS data, as its strengths lie in the identification of patterns in high-dimensional input data.
In addition, deep learning models are well-suited for execution on accelerators, highly parallelized computing units such as graphics processing units (GPUs), helping to address the computational challenges associated to the large data volumes.
Deep learning models are already well established for processing waveforms from classical seismic instruments \citep{zhuPhaseNetDeepneuralnetworkbasedSeismic2019,mousaviEarthquakeTransformerAttentive2020,munchmeyerTransformerEarthquakeAlerting2020}.
For DAS, deep learning models have been proposed for different tasks, such as seismic phase picking \citep{zhuSeismicArrivaltimePicking2023,xiaoDeepSubDASEarthquakePhase2026}, denoising \citep{vandenendeSelfSupervisedDeepLearning2023,lapinsDASN2NMachineLearning2024,zittSelfSupervisedCoherenceBasedDenoising2025}, and
event classification of anthropogenic events \citep{tomasovComprehensiveDatasetEvent2025}.
Even full event monitoring workflows, involving phase picking, phase association, and earthquake location, have been presented \citep{bailletAutomaticEarthquakeCatalogs2025}.

However, currently there is a lack of standardization among the proposed approaches.
Different models use different interfaces, reducing interoperability and comparability.
For example, it is difficult for practitioners to test different models on their data, as each model provides a different API and different requirements.
In addition, implementations are often tied to the data formats of specific interrogators, requiring manual data conversion.
Similarly, there exists no standard for benchmark datasets that are integral for training and evaluating deep learning models.
These challenges closely mirror the early days of deep learning for classical seismic data \citep{munchmeyerWhichPickerFits2022,woollamSeisBenchToolboxMachine2022}.

Here, we present SeisBench DAS, a framework for machine learning on DAS data.
SeisBench DAS extends the existing SeisBench framework \citep{woollamSeisBenchToolboxMachine2022} with functionality for DAS models and datasets.
The central focus of SeisBench DAS is to provide simple, unified application programming interfaces (APIs) for using deep learning models for DAS and accessing benchmark datasets.
To this end, SeisBench DAS defines a data format for DAS benchmark datasets, an API to access these datasets, and a unified model API.
To resolve the dependency on specific input data formats, SeisBench DAS delegates the input to the xdas framework \citep{trabattoniXdasPythonFramework2025}, offering read capabilities for a wide range of data format from all common interrogators.
SeisBench DAS closely follows the design principles of SeisBench and shares the core code base.
The framework is designed to be extensible with new models and datasets, building on the community around SeisBench.
In the following, we provide a brief overview of SeisBench for classical data, introduce the novel challenges related to DAS data, discuss the details of SeisBench DAS, and present examples for the usage of the framework.

\section{The SeisBench framework}

\begin{figure*}[ht!]
	\centering
	\includegraphics[width=\textwidth]{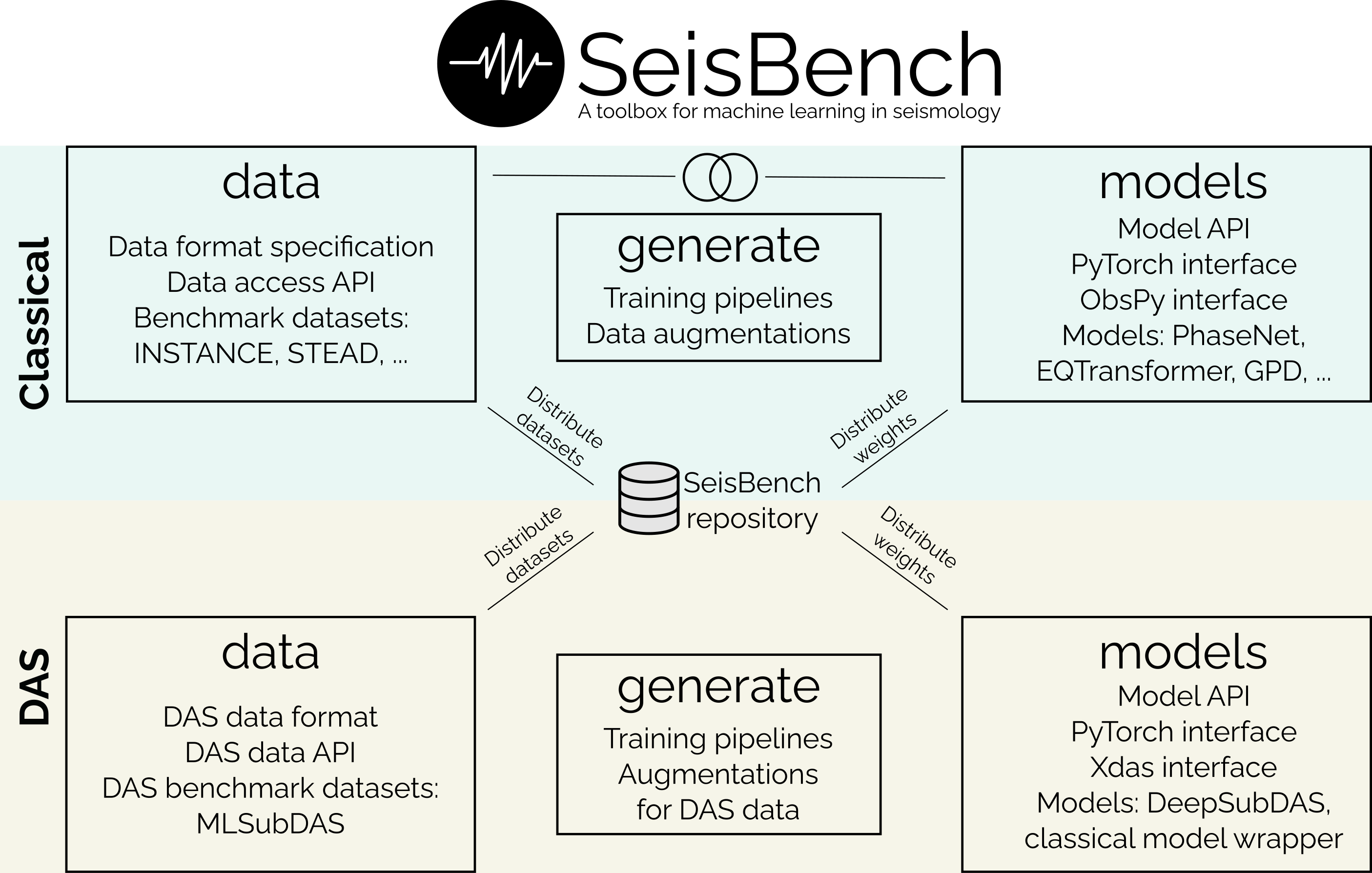}
	\caption{Overview of the SeisBench framework with the three core components: data, generate, and models. The top half (green) shows the classical components of SeisBench. The bottom half (yellow) shows the novel extensions for DAS data.}
	\label{fig:overview}
\end{figure*}

\emph{SeisBench: A toolbox for machine learning in seismology} (short SeisBench) is an open-source Python framework \citep{woollamSeisBenchToolboxMachine2022}.
SeisBench combines the different aspects of machine learning for seismic data with a focus on applicability, comparability and reproducibility.
It integrates datasets and models in unified formats, aiming to be accessible to both model developers and seismological practitioners.
It consists of three core components (Figure~\ref{fig:overview}, green part): \emph{seisbench.data}, \emph{seisbench.models}, and \emph{seisbench.generate}.
We provide an overview of the key functionalities of each module below, before discussing how we integrate the processing of DAS data into the modules.

\emph{seisbench.data} contains all functionality related to creating, managing and using benchmark datasets.
At the core of the module is a data format definition for benchmark datasets.
Each dataset consists of a CSV file with metadata and an HDF5 file with waveforms.
The metadata columns follow a consistent naming scheme, but are flexible with respect to additional bespoke columns.
Each line in the metadata corresponds to one multi-component trace of seismic data.
The format is deliberately simple to allow users to build and access these datasets without necessarily using SeisBench.
As of August 2026, more than 25 datsets in this format are part of SeisBench, for example, INSTANCE \citep{micheliniINSTANCEItalianSeismic2021}, STEAD \citep{mousaviSTanfordEArthquakeDataset2019}, and PNW \citep{niCuratedPacificNorthwest2023}.
SeisBench provides a Python interface for accessing data in this format and distributes datasets through the SeisBench repository, hosted on the high-performance dCache infrastructure of DESY \citep{fuhrmannDCacheStorageSystem2006}.

\emph{seisbench.models} provides access to deep learning models for seismic data processing.
Each model is built using PyTorch \citep{anselPyTorch2Faster2024}, but provides two distinct interfaces.
First, a standard PyTorch interface makes it possible to train the models like any other PyTorch model using typical frameworks.
However, this interface is not convenient for practitioners with data in, for example, mseed or sac format.
Therefore, SeisBench models also implement an Obspy interface \citep{beyreutherObsPyPythonToolbox2010}.
The \emph{annotate} and \emph{classify} functions take an ObsPy \emph{Stream} object and return either probability time series or discrete objects, such as phase picks.
Internally, SeisBench prepares the \emph{Stream} for annotation with PyTorch, for example, by resampling, aligning components, and chunking, passes it to the PyTorch interface, and converts the outputs back.
Combined with ObsPy's capability to read most common seismic data formats, this makes it straightforward to apply SeisBench to waveform data.
Both interfaces can be executed on one or more CPUs or any accelerators, such as GPU, XPU, or MPS, with a single line of code.

SeisBench implements a number of different model architectures, such as PhaseNet \citep{zhuPhaseNetDeepneuralnetworkbasedSeismic2019}, EQTransformer \citep{mousaviEarthquakeTransformerAttentive2020}, and PickBlue \citep{bornsteinPickBlueSeismicPhase2024}.
SeisBench complements these with a large set ($>80$) of pretrained model weights that are distributed through the SeisBench repository, covering different seismogenic regions and data types.
By linking the interfaces for training and for application, and by providing an easy way to access pretrained weights, SeisBench aims to close the gap between model developers and seismological practitioners.

\emph{seisbench.generate} links data and models through training pipelines.
SeisBench training pipelines consist of a sequence of augmentations that transform the samples in the dataset into samples that can be passed to the models.
Such transformations are necessary, as the data in the benchmark datasets is often not directly compatible with the models, for example, due to input length requirements.
Furthermore, data augmentations can increase the diversity of the training data and thus result in more robust models.
For efficient model training and evaluation, the training pipelines are tightly integrated with PyTorch utilities.

Since its original release, SeisBench has been substantially improved and extended through contributions from the community and the maintainers.
These include bugfixes, stability and performance enhancements, and feature additions.\footnote{For a complete overview, please refer to the release history (\url{https://github.com/seisbench/seisbench/releases}) or the list of pull requests on Github (\url{https://github.com/seisbench/seisbench/pulls}).}
In particular, a wide range of datasets and models have been added, covering diverse seismogenic regions, such as tectonic, volcanic, and induced seismicity, and a range of seismological tasks, such as phase picking, waveform denoising, and earthquake depth estimation.

\section{Requirements of DAS processing}

The different nature of DAS and classical waveform data leads to different requirements that have guided the design of SeisBench DAS.
In particular, we took into account differences in dataset size, compute requirements, and data access patterns.
In terms of data size, DAS data is orders of magnitude bigger than classical seismic data.
While 1~TB is already a noteworthy amount of data for classical waveforms, it is a small dataset for DAS standards.
As a corollary, while for classical seismic data larger-than-memory processing is not a primary concern, for DAS data it will be relevant for almost all cases.
In addition, iterating over days and stations is trivial to implement for classical data and even allows easy parallelization, while for DAS a similar chunking needs to be much more fine-grained and properly handle all overlaps.
For this reason, the SeisBench implementation for DAS puts a focus on larger-than-memory processing, while it is not part of the original SeisBench.

As a direct consequence of the data size, the compute requirements for processing DAS data are substantially higher, especially for deep learning models, which are considerably heavier for DAS data than classical waveforms.
At the same time, the data structure of DAS, consisting of large matrices, is well-suited for accelerators such as GPUs.
In particular, a chunk of DAS data is big enough to efficiently use the full compute units of a GPU, while for classical data typical chunks only use a fraction of the GPU.
While for classical data processing on GPU is often not economical \citep{kraussSeismologyCloudGuidance2023} and even petabyte-scale workflows can be executed efficiently on CPU \citep{niGlobalscaleDatabaseSeismic2025,niReviewCloudComputing2025}, similar processing for DAS requires accelerators.
Therefore, efficient processing and accelerator support throughout all processing steps, while implemented in the classical SeisBench, takes a more central role for SeisBench DAS.

While the larger data size of DAS records introduces the challenges outlined above, the highly regular structure also enables significant efficiency gains.
In particular, the read and write performance is much easier to optimize.
For classical data, the individual data, for example, a day of waveforms in mseed format with STEIM2 encoding, is typically on the scale of a few megabytes.
Read throughput is therefore often controlled by the high number of random accesses, frequently making  IO the bottleneck.
In contrast, even short DAS segment sizes typically exceed 100~MB, allowing for efficient sequential reads.
Furthermore, while classical data often uses seismology-specific formats and encodings, most DAS formats rely on the widely-used hdf5 format without compression.
For these formats, standard libraries implement memory mapping through virtual arrays and thus enable larger-than-memory processing without requiring excessive custom code for data management.

\section{SeisBench DAS integration}

For the integration of DAS into SeisBench, we follow a similar structure as for the classical waveform data.
At the same time, we take into account the characteristics outlined above and lessons learned from the ongoing development and maintenance of SeisBench.
SeisBench DAS adds components to the \emph{data}, \emph{models}, and \emph{generate} submodules of SeisBench, extending the underlying concepts to DAS data (Figure~\ref{fig:overview}, yellow part).
SeisBench DAS is distributed as part of the SeisBench python package.

\subsection{Benchmark datasets}

\begin{figure*}[ht!]
	\centering
	\includegraphics[width=\textwidth]{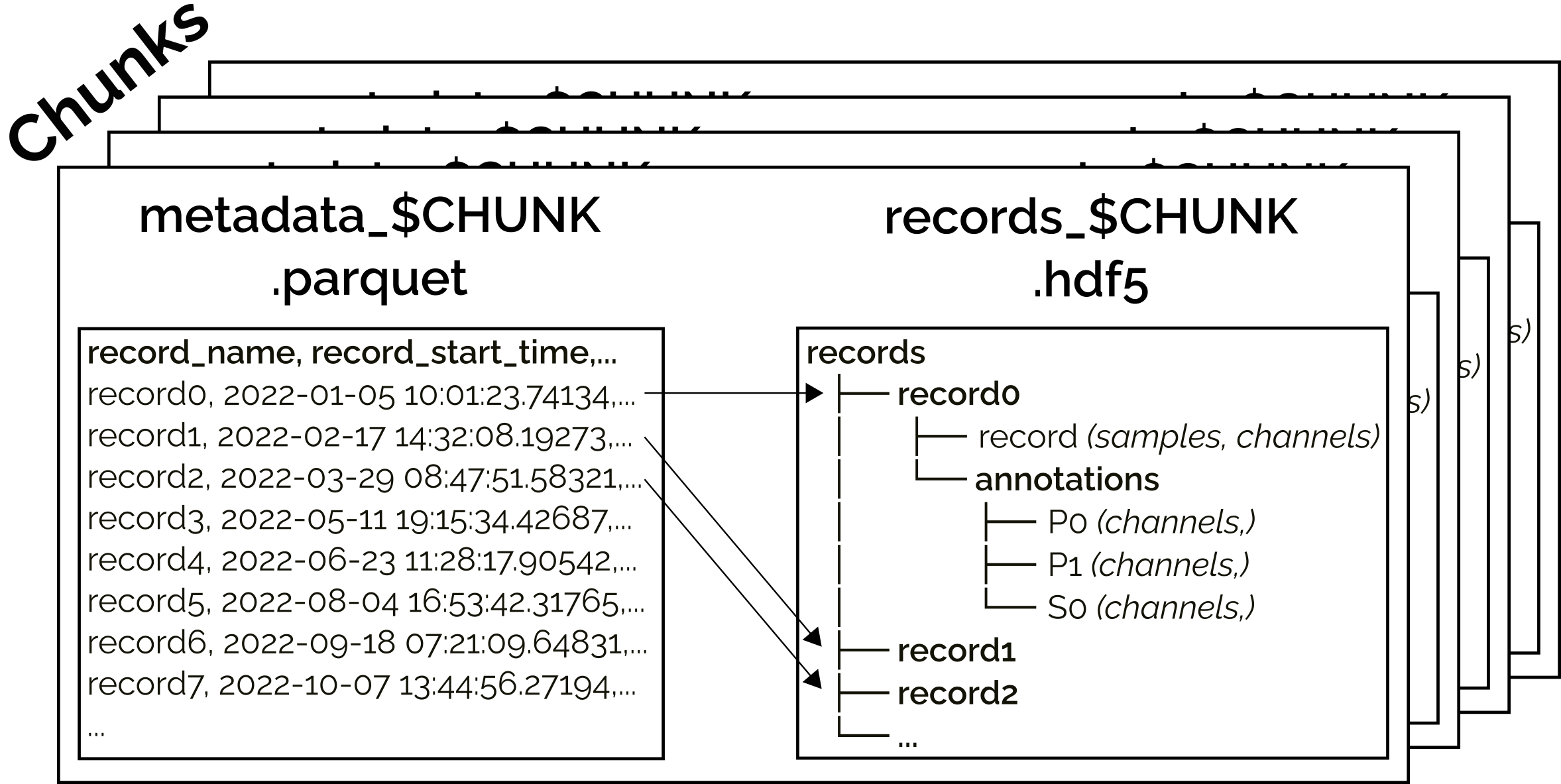}
	\label{fig:data_format}
	\caption{Visualization of the data format, consisting of a metadata file and a records file. Arrows indicate how metadata entries in the parquet file point to associated records and annotations in the hdf5 file. Italic entries in brackets indicate array shapes.}
\end{figure*}

\begin{figure*}[ht!]
	\centering
	\includegraphics[width=\textwidth]{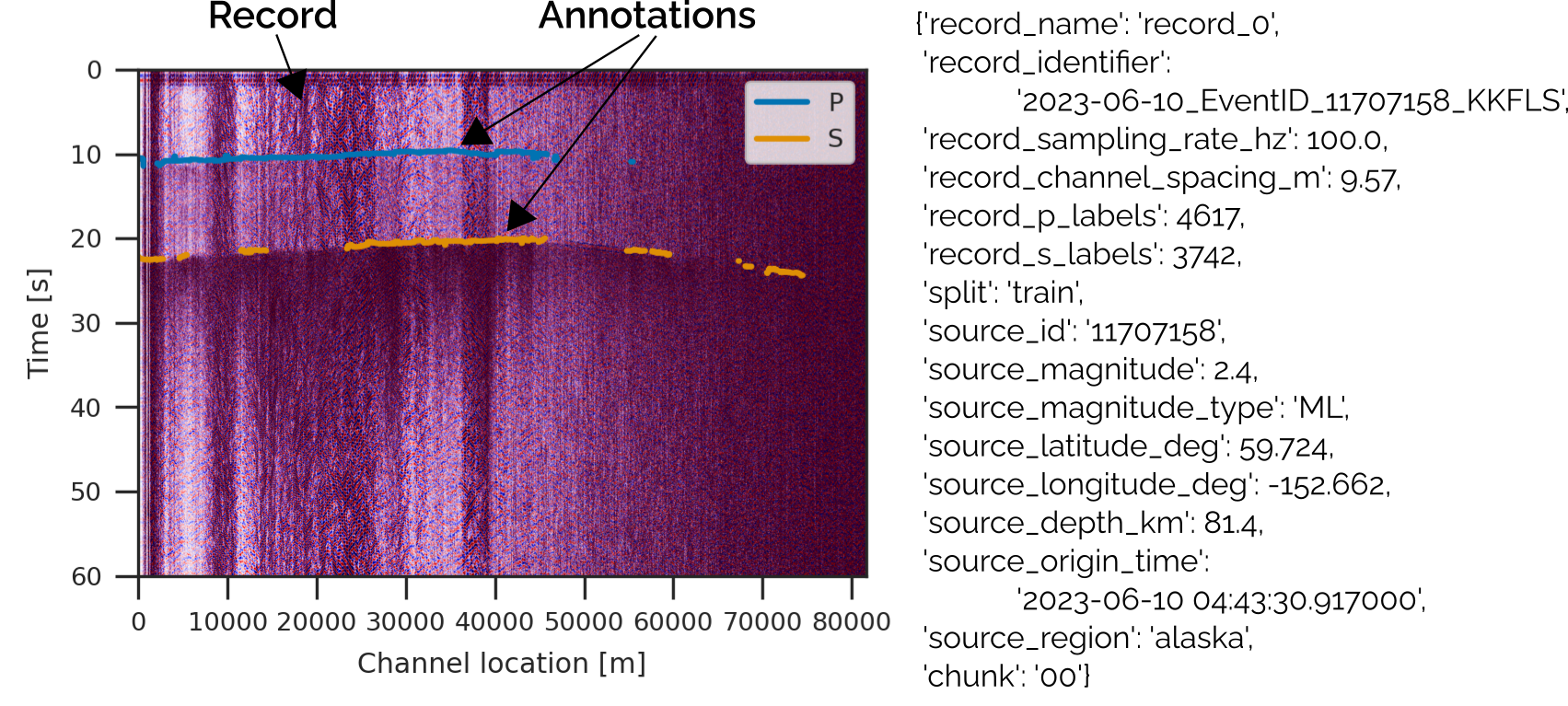}
	\caption{An example record from the MLSubDAS dataset, showing the record, the annotations, and the metadata. Note that the `chunk` keyword is not part of the metadata on disk but added by SeisBench in-memory.}
	\label{fig:example_record}
\end{figure*}

SeisBench DAS provides a data format for DAS datasets and an API to interact with datasets in this format.
This is complemented by the established functionality of distributing benchmark datasets through the SeisBench repository.
As for classical datasets, each dataset consists of a metadata file and a hdf5 file containing the actual records.
We use the same metadata naming scheme as for classical data, but rename the category ``trace'' to ``record''.
For the metadata, we changed the format from csv to parquet \citep{apacheparquetformat}.
Compared to csv files, parquet files are typed, more space-efficient, and offer faster read-write performance.
While parquet files are not human-readable, they are supported by libraries in all common programming languages.
Each row in the metadata is linked to a DAS record, stored in the hdf5 file.
Each record is a two-dimensional array, where the first dimension indexes samples and the second dimension indexes channels.

In addition to the metadata, each record can be associated with a set of annotations.
An annotation is a 1D array with the same length as the number of channels for each record.
This way, annotations can store channel-specific information, for example, the arrival sample of P waves.
Channels without the features, for example, without a labeled arrival, are indicated by NaN values.
The annotations are stored in the same hdf5 file as the records.
Each record can have a different set of annotations.
For example, some records might have P and S wave annotations, while others have only P waves, only S waves, no arrivals at all, or even multiple P and S waves.
Figure~\ref{fig:example_record} shows an example records with metadata and annotations.

For DAS benchmark datasets, we follow an index-based view.
This means, that the record is an array with equal spacing in time and space dimension.
While the start time of the record, the sampling rate, and the channel spacing should be indicated in the metadata, no further coordinate-awareness exists.
This stands in contrast to coordinate-aware libraries, such as xdas, where arrays can be accessed and sliced based on, for example, time or distance ranges.
For the benchmark datasets, we chose this index-based view, because it is the natural input for deep learning models, in particular, convolutional neural networks that can not be applied to non-uniform sampling in space or time.

SeisBench provides an API to access benchmark DAS datasets, largely following the API for classical waveform datasets.
However, there are a few key differences.
First, in addition to metadata and waveforms, the dataset provides access to all annotations for a record.
Second, the dataset API does not provide an option for automatic data resampling.
While this is a useful function for classical datasets, it often leads to unexpected behavior and substantial CPU usage in a supposedly cheap data access call.
Instead, resampling should be done explicitly during the training or evaluation process if required.
Third, instead of returning in-memory arrays with the data, by default, the API only returns virtual arrays, i.e., pointers to the record within the hdf5 file.
This avoids loading the data into memory, instead performing only a simple index lookup.
This is useful, as often only part of the data needs to be loaded.
Control over the data loading is passed to the downstream process.
As individual records are much bigger than waveform traces in classical datasets, SeisBench does not implement caching for DAS data access but rather relies on the underlying hdf5 library and file system.

Due to the size of individual DAS records, even datasets with only a few hundred records can be hundreds of GB in size.
To reduce complications in data transfer and management, SeisBench implements an optional chunking mechanisms for DAS datasets.
To this end, the dataset is split into individual metadata and data files, with each pair itself a valid DAS dataset.
The total set of available chunks is listed in an associated \emph{chunks} file.
From the user perspective, the access API is identical for chunked and non-chunked datasets.

\subsection{Model interface}

\begin{figure*}[ht!]
	\centering
	\includegraphics[width=\textwidth]{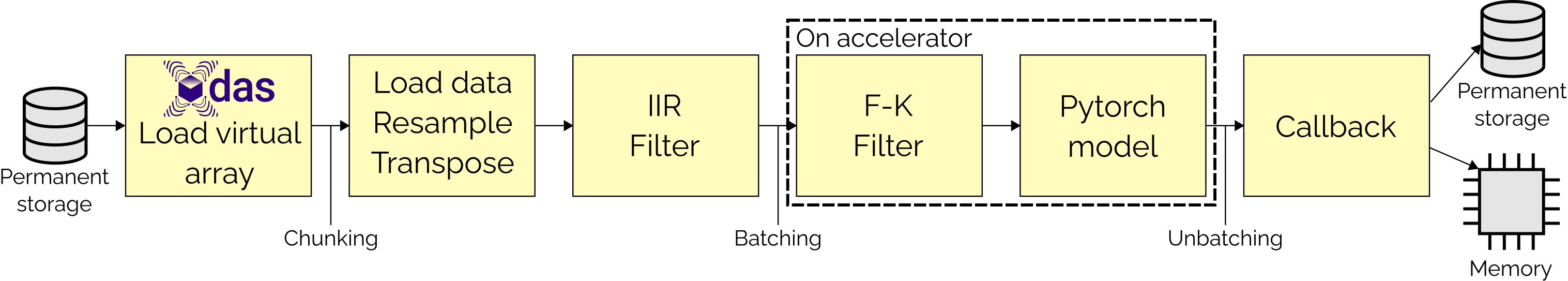}
	\caption{Schematic overview of the steps in the annotation pipeline of SeisBench DAS. The dashed box indicates the steps that can be performed on accelerators, such as GPUs.}
	\label{fig:annotate}
\end{figure*}

The SeisBench DAS model API is closely inspired by the one for classical seismic waveforms.
Each model exposes two interfaces.
The first interface is the PyTorch interface and mostly intended for model training.
The second interface is targeted at practitioners seeking to apply the models and is accessible through the \textit{annotate} and \textit{classify} functions.
As the first interface is standard PyTorch, we do not provide a further description here.
Instead, we provide an extended description of the interface for practitioners, starting with the frontend and moving to the backend afterwards.
An overview of this pipeline is shown in Figure~\ref{fig:annotate}.

For managing file input and output, SeisBench DAS uses xdas, a Python framework for distributed Acoustic Sensing \citep{trabattoniXdasPythonFramework2025}.
xdas enables reading diverse DAS data formats and provides an uniform interface.
This way, the IO in xdas becomes decoupled from the processing of the data in SeisBench.
In addition, xdas supports virtual datasets, enabling larger-than-memory processing.

SeisBench models for classical waveforms return either full characteristic functions (\emph{annotate}) or discrete objects like picks (\emph{classify}).
For DAS processing, we follow the same principle.
However, full arrays with annotations typically have a similar size to the input data and therefore will often not fit into memory.
To address this challenge, we introduce callbacks.
Each annotation that has been returned from the model is passed to the callback and afterwards evicted from memory, keeping the total memory almost constant.
Currently, callbacks for collecting the whole annotation in memory, saving it to disk, and phase picking are available.
Users can either choose to provide their own callback (\emph{annotate}) or use the suggested callback of the model (\emph{classify}).
For example, phase picking models will use the phase picking callback by default.

As full datasets are often too large to keep in memory, SeisBench DAS processes data in chunks.
As input, we rely on virtual datasets from xdas, providing an interface to load targeted chunks from the disk.
The chunk size is determined by the model, either using a fixed size or deriving the size from the shape of the input data.
In addition, the model defines an overlap between adjacent chunks.
SeisBench then iterates over the input chunks, preprocesses them, passes them to the deep learning model, and passes the outputs to the callback.
The implementation is asynchronous, prefetching the next data chunk while the model is processing.
In addition, it uses batching for the deep learning model to improve compute efficiency, in particular, on GPUs.

For classical models, SeisBench implements automatic resampling to a model-defined sampling rate.
However, DAS models are often trained on a mixture of different sampling rates and channel spacings.
In addition, classical models have proven robust against changes in sampling rate between training and application \citep{shiLabquakesMegathrustsScaling2024}.
Therefore, SeisBench DAS models can specify an acceptable range of channel spacings and sampling rates.
If the input values are outside this range, SeisBench will automatically be resample the input.
Resampling is applied independently to each chunk on-the-fly and uses a zero-phase FIR filter to avoid alias artifacts.
Boundary artifacts are avoided by loading additional samples on each side of the input data.

For preprocessing, SeisBench provides the opportunity to apply a filter along the time dimension or an F-K filter.
For filtering along the time axis, all IIR filters available in \textit{scipy.signal} are available \citep{virtanenSciPy10Fundamental2020}.
Only causal filters are supported, as zero-phase filters would require to preload the full dataset.
SeisBench automatically keeps track of filter states across different chunks to ensure that chunk boundaries do not introduce artifacts.
F-K filters have turned out to be highly effective at noise suppression for DAS data \citep[e.g.][]{iskenDenoisingDistributedAcoustic2022}.
The F-K filter is implemented in PyTorch and thus can be applied to batched data, making its application highly efficient.
The F-K filter can introduce boundary artifacts, but it is not possible to apply it to the whole data at once without loading it into memory completely.

Two callbacks, the one for in-memory collection and the writer callback need to splice the chunked data back together into one array.
To avoid discontinuities at the boundaries between chunks, SeisBench uses a smooth interpolation.
Each output pixel is weighted using a windowing function and pixels in the overlap are averaged using the weighted mean.
For the writer, the splicing logic is designed to minimize the amount of data that has to be kept in memory.
Once a whole row of the data has been processed, it is written to disk and evicted from memory.
The writer keeps tracks of overlaps and only writes output data once the annotations from all windows containing a pixel have been collected.

For compute efficiency, SeisBench DAS supports accelerators, such as GPUs.
All accelerators supported by PyTorch can be used.
SeisBench moves data to the accelerator as early as possible, directly after potential resampling and IIR filtering.
The F-K filter, if active, is executed on the accelerator, avoiding the cost of the Fourier transform on CPU.
Annotations are moved back to the CPU before passing them to the callback.

While several models specific to DAS have been published by now, there is a far larger set of established models for classical waveform data.
Therefore, SeisBench DAS also implements a wrapper to apply these models to DAS data.
The wrapper transforms the single channel DAS data to three-component data either by putting the same record multiple times or padding the missing channels with zeros.
Preprocessing arguments from the classical model are automatically translated to the corresponding arguments for the DAS model.
The wrapper can be used around all sequence-to-sequence models available in SeisBench, for example, PhaseNet or EQTransformer.
The wrapper uses the identical interface to the DAS-native models.

\subsection{Training pipelines}

To connect benchmark datasets with DAS models, we implemented flexible and extensible training pipelines.
Such pipelines are necessary, as the data format available in the benchmark datasets might not exactly match the format required as input to the models.
For example, models typically require specific input shapes, while the samples in the dataset can have varying shape.
In addition, the labels need to be encoded in an appropriate format for the model, for example, as probability curves when they are available in a different form, e.g., a list of discrete pick times for each channel.
Furthermore, augmentations can be used to generate different training samples from the same sample in the dataset, enhancing the variety of training data and thereby often improving training effectiveness.
These pipelines have been integrated into the \emph{seisbench.generate} module.

Training pipelines consist of a sequence of augmentations that transform the data in a predefined way.
All augmentations work on representations of the data as \emph{numpy} arrays or virtual \emph{hdf5} datasets.
As of now, SeisBench DAS offers augmentations for window selection and data labelling.
By default, window augmentations take virtual arrays as input and output in-memory arrays.
This way, only the data actually passed to the model needs to be read into memory.

In contrast to the generation pipelines for classical waveform data, SeisBench DAS offers no augmentations for data normalization.
Instead, data normalization should be included in the \emph{forward} step of the PyTorch model.
This pattern ensures consistency in normalization between the PyTorch and xdas interfaces.
Such inconsistencies have previously led to degraded performance with classical models in SeisBench.\footnote{For details, see \url{https://github.com/seisbench/seisbench/issues/187} and \url{https://github.com/seisbench/seisbench/pull/188}.}
In addition, including the normalization in the \emph{forward} step increases the computational efficiency, as it is then performed in PyTorch, allowing for batch processing and the use of accelerators, such as GPUs.
Similarly, for F-K filtering, we refer to the F-K filter PyTorch module implemented in SeisBench DAS.

As for classical waveform data, SeisBench DAS training pipelines are tightly integrated with PyTorch datasets.
The generation pipeline inherits from the PyTorch \emph{dataset} class, allowing a direct use in PyTorch data loaders and thus enabling parallel processing.
The SeisBench interface ensures that the underlying datasets can be seamlessly digested in this multiprocessing setup.

\section{Example application}

\subsection{Applying a model}

As a first example, we present a workflow for applying a phase picking model to continuous DAS data (Listing~\ref{lst:picking}).
For the example, we use the DAS-native DeepSubDAS model.
The steps for model application are: loading the model with a pretrained set of weights, moving the model to an accelerator if available, loading the data with xdas, defining the callback, and applying the model.
The output of the model is available as a pandas data frame with the columns ``time'', ``confidence'', ``channel'', and ``phase''.
The same workflow can be run with different models, for example, wrapped models for classical seismic data, by changing the model instantiation call.

\begin{lstlisting}[caption={Example application of DeepSubDAS for phase picking with SeisBench. The DASPickingCallback sets the confidence thresholds when a pick is declared; the smaller that number the more picks will be declared, reducing the number of missed picks, but increasing the number of erroneous picks. The batch size controls the trade-off between computation speed and memory use, with larger batch sizes favouring faster execution at the cost of higher memory consumption.}, label=lst:picking, language=Python]
import seisbench.models as sbm
import xdas

model = sbm.DeepSubDAS.from_pretrained("original") # Load weights from cache or repository
model.to_preferred_device()                        # Move model to accelerator if available
data = xdas.open("das_example.hdf5")               # Load metadata and create virtual array
callback = sbm.DASPickingCallback(thresholds={"P": 0.2, "S": 0.2})  # Define call back

model.annotate(data, callback, batch_size=4)       # Perform the computation

picks = callback.get_results_dataframe()           # Return pandas data frame with picks
\end{lstlisting}

\subsection{Training a model}

As a second example, we show a workflow for training a DeepSubDAS model on the MLSubDAS dataset (Listing~\ref{lst:training}).
This workflow combines all three parts of SeisBench: \emph{models}, \emph{data}, and \emph{generate}.
For the example, we use a random window with a fixed size of 1000 samples and 500 channels.
Each time a sample from the dataset is used, a different window is selected at random, while guaranteeing that the window contains at least one annotation.
For encoding the labels, we use probabilistic labels along the sample axis, similar to the ones introduced for PhaseNet \citep{zhuPhaseNetDeepneuralnetworkbasedSeismic2019}.
Accordingly, we use a cross-entropy loss for scoring the models.
We optimize the model with the Adam optimizer.
For brevity, we avoid monitoring a validation loss or evaluating the final model.

\begin{lstlisting}[caption={Example training workflow for DeepSubDAS with SeisBench. The examples shows a minimal functional workflow and does not contain aspects such as monitoring validation loss, output logging, or parallelization in data loading.}, label=lst:training, language=Python]
import seisbench.data as sbd
import seisbench.models as sbm
import seisbench.generate as sbg
import torch
import torch.nn.functional as F

# Load metadata of training data. If the data is not available locally, it will be downloaded
# from the SeisBench repository. Create virtual array
data = sbd.MLSubDAS()

# Create model with pretrained ResNet101 weights and move it to accelerator if available
model = sbm.DeepSubDAS(deeplab_weights_backbone=ResNet101_Weights.IMAGENET1K_V1)
model.to_preferred_device()

# Define augmentations.
window = sbg.RandomDASWindow(				# select window
    shape=(1000, 500),
    contains_annotation=True,
)
labeller = sbg.ProbabilisticDASLabeller(    # select labelling function
    annotation_mapping={
        **{f"P_{i}": "P" for i in range(100)}, # Annotation keys are flexible regarding
        **{f"S_{i}": "S" for i in range(100)}, # the number of arrivals per record.
    },
    noise_map=True,
)

# Add augmentations to sample generator
# The order of adding the augmentations defines the order of execution
train_generator = sbg.DASGenerator(data.train())
train_generator.augmentation(window)
train_generator.augmentation(labeller)

# Create data loader
train_loader = DataLoader(
    train_generator,
    batch_size=8,
    shuffle=True,
)

# Create optimizer
optimizer = torch.optim.Adam(model.parameters(), lr=1e-3)

# Define loss function
def fuzzy_cross_entropy_loss(logits, target_probs):
    log_probs = F.log_softmax(logits, dim=1)
    loss = (-target_probs * log_probs).sum(dim=1)
    loss = loss.mean()
    return loss

# Training loop
for epoch in range(100):
    for batch_id, batch in enumerate(dataloader):
        # Compute prediction and loss
        x = batch["X"].to(model.device)
        pred = model(x)
        loss = fuzzy_cross_entropy_loss(pred["full"], batch["y"].to(model.device))

        # Backpropagation
        optimizer.zero_grad()
        loss.backward()
        optimizer.step()

\end{lstlisting}

\subsection{Performance evaluation}

\begin{figure*}[ht!]
	\centering
	\includegraphics[width=\textwidth]{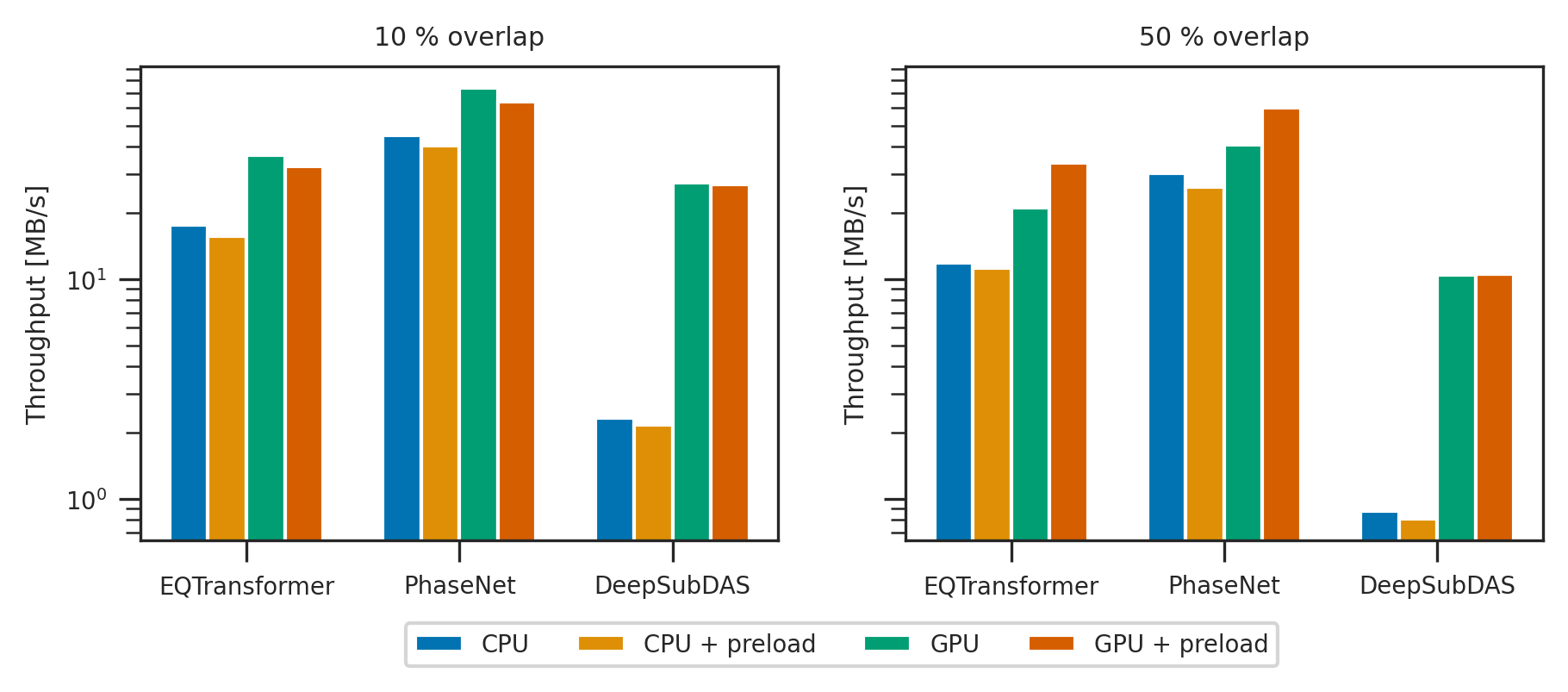}
	\caption{Annotation throughput for different models and configurations. Models are executed either on CPU or on GPU. Models are either applied to data stored on a local SSD or to data preloaded into memory (preload). The columns differ in the overlap between adjacent windows. For EQTransformer and PhaseNet, overlap is only used along the sample axis. For DeepSubDAS, the same overlap is used along both axes.}
	\label{fig:performance}
\end{figure*}

We evaluate the annotation performance of SeisBench DAS in different configurations.
As input, we use a 3.1~GB continuous DAS record with 36,718 channels and 225~s long records at a sampling rate of 100~Hz.
The data type are 32~bit floats.
We test three models, two wrapped models for classical waveforms (EQTransformer, PhaseNet) and one DAS-native model (DeepSubDAS).
For the wrapped models, we clone the DAS record into all 3 components.
We use overlaps of 10~\% and 50~\% between adjacent applications of the deep learning models along the sample axis for all models.
Along the channel axis, we only use overlap for DeepSubDAS as the wrapped models are independent between channels.

We test four configurations of the model, differing in the compute infrastructure and data management.
We test each model on CPU and on GPU.
Each run uses 16 cores of an Intel Xeon Platinum 8360Y CPU and 64~GB of main memory.
GPU runs have exclusive access to one Nvidia A40 GPU.
In addition, we run all tests once with data preloaded into memory and once with data on a local SSD.
We calculate raw throughput values, defined as the file size divided by the total run time.
We only count the run time for the call to \emph{annotate} and not any potential overheads, such as the initial model loading or the CUDA intitialization.

We observe annotation throughputs between 0.8~MB/s and 74~MB/s.
For the wrapped classical models, CPU speeds are between 11 and 17~MB/s for EQTransformer and between 26 to 45~MB/s for PhaseNet.
GPU speeds are approximately 2x faster with a similar difference between PhaseNet and EQTransformer.
The lower throughput for EQTransformer results from the increased compute requirement due to the higher model complexity.
For DeepSubDAS, the difference between CPU and GPU is much more pronounced, with a factor of roughly 13 between 0.8 to 2.3~MB/s on CPU and 10 to 27~MB/s on GPU.
Notably, even on GPU DeepSubDAS only achieves lower or similar throughput compared to the wrapped classical models on CPU.
This highlights the substantially larger compute requirements of the DAS-native model.
Throughputs are in all cases higher when using lower overlaps, a natural consequence of fewer model applications.
This difference is more pronounced for DeepSubDAS as overlap is applied along both axes.
As expected based on the asynchronous implementation, the difference in annotation speeds between preloaded and non-preloaded runs is small.

To compare the runtimes to the length of the record, we calculate how many seconds of data each model can annotate per second of compute time.
On GPU, we reach 4.3~s/5.3~s for PhaseNet, 2.4~s/2.6~s for EQTransformer, and 0.75~s/2.0~s for DeepSubDAS (10~\%/50~\% overlap).
On CPU, we reach 2.2~s/3.2~s for PhaseNet, 0.8~s/1.3~s for EQTransformer, and 0.06~s/0.17~s for DeepSubDAS.
This mean that on a single GPU, only the wrapped models or DeepSubDAS with low overlap are ready for near-real-time application.
In addition, for batch processing, a full annotation on a single GPU will take at least a fifth of the record duration.

To improve runtimes, several approaches are feasible.
First, a low overlap avoids duplicate computations.
However, models need to to be trained accordingly to minimize boundary artifacts.
Second, at more than 36,000 channels, our example uses a very long cable.
Most deployments will not have as many usable channels.
In addition, information is often redundant between channels and processing a subset will produce almost identical results.
Picking on a reduced number of channels will boost performance, as the run time is almost linear in the number of channels.
Third, for large processing, parallelization is inevitable.
For example, the models can be applied to daily chunks of data, giving near-linear speed-ups with very little boundary artifacts.
Nonetheless, these results demonstrate that while for classical seismic data, compute requirements are no relevant limitation for phase picking, for DAS data efficient models and implementations are essential.
Therefore, it will also be worthwhile developing DAS-native picking models with optimized trade-offs between runtime and picking performance.

\section{Conclusion \& Outlook}

In this paper, we presented SeisBench DAS, which adds functionality for distributed acoustic sensing data to the established SeisBench framework.
SeisBench DAS adresses the need for standardization in deep learning models for DAS data, as well as in DAS benchmark datasets.
In addition, SeisBench DAS implements a data processing engine to directly apply deep learning models implemented in PyTorch to DAS data in diverse formats, using the xdas framework for data access.
The engine is designed for larger-than-memory processing, taking into account the typical requirements posed by DAS data.
As an example, we implemented the MLSubDAS dataset, the DeepSubDAS model, and a wrapper to efficiently apply models for classical seismic data to DAS records.

While SeisBench DAS introduces a metadata naming and data format for DAS benchmark datasets, this is not intended as a wider standard for general purpose.
The proposed format is geared towards machine learning and lacks sufficient generalizability for general purpose approaches.
To this end, we refer to ongoing efforts for the standardization of DAS data \citep{quinteros2026modern} and metadata \citep{lai2024toward}.
Through the integration of SeisBench DAS with xdas, we ensure that SeisBench will be compatible with these future data standards.

For an easy introduction to SeisBench DAS, we provide tutorials in the form of Jupyter notebooks.
The tutorials can be run locally, but also on Google Colab with a single click.
In addition, a full documentation for SeisBench and SeisBench DAS is hosted at \url{https://seisbench.readthedocs.io/}.
As the DAS functionality substantially extends the SeisBench framework, we have restructured the whole SeisBench documentation and the tutorial overview for easier navigation.
To ensure high code quality, SeisBench DAS follows the development principles of SeisBench, for example, version management through Git, mandatory code formatting and linting, and unit tests with continuous integration.
Community interaction is facilitated through Github issues and pull requests.

While SeisBench DAS integrates some models and datasets, the initial collection is far from comprehensive.
For example, the initially included models and dataset only cover specific use cases.
In addition, as shown in the performance evaluation, there is a requirement for more compute-efficient DAS-native models.
The contribution of this initial version of SeisBench DAS is therefore less the specific models and datasets included, but rather to propose a standard for models and datasets and to provide a processing engine connecting PyTorch models directly to DAS data through xdas.
With the extensible structure of SeisBench DAS, we aim to grow a diverse framework for machine learning on fibre-optic data.

\begin{acknowledgements}
We gratefully acknowledge the support of EU-INFRATECH, Grant agreement ID: 101095055 (SUBMERSE project).
This work utilized high-performance computing resources made possible by funding from the Ministry of Science, Research and Culture of the State of Brandenburg (MWFK) and is operated by the IT Services and Operations unit of the Helmholtz Centre Potsdam.
We acknowledge the Helmholtz Association for providing data storage and distribution capabilities through the dCache service hosted at DESY.
We thank Alister Trabattoni for support with using the xdas library.
JM acknowledges his independent research fellowship (GFZ Discovery Fellowship).
\end{acknowledgements}

\section*{Data and code availability}
SeisBench is available through PyPI (\texttt{pip install seisbench[das]}), on Github (\url{https://github.com/seisbench/seisbench}), and Zenodo (\url{http://doi.org/10.5281/zenodo.5568812}).
Full DAS support is available from SeisBench v0.12 onwards.
SeisBench is available under the open GPL-3.0 license.

\section*{Competing interests}
The authors declare no competing interests.

\bibliography{Zotero,mybibfile}

\end{document}